\documentstyle[multicol,aps,epsf]{revtex}
\begin{document}
\draft
\title{``Splitting'' of Delocalized States  in a Double--Layer 
System in a Strong Magnetic Field}

\author{A. Gramada\cite{pos}
 and M. E. Raikh}
\address{Department of Physics, University of Utah, Salt Lake City, 
Utah  84112}

\maketitle

\widetext

\begin{abstract}
A double--layer system in a strong perpendicular magnetic field
is considered. We assume a random potential in each layer to be  smooth. 
We also assume that there is no correlation between random potentials in 
different layers. Under these conditions the equipotential lines from 
different layers, corresponding to the same energy, may  cross each other. 
We show that, if the tunnel coupling between the layers exceeds some 
characteristic value (which is much smaller than the width of the Landau 
level), then  the probability for an  electron to switch equipotential 
(and, thus, the layer) at the intersection is close to one.     
As a result, the structure of each delocalized state in a double--layer
system becomes completely different from that for an isolated layer. 
The state is composed of alternating pieces of equipotentials from 
different planes. These combined equipotentials form a percolation network. 
We demonstrate that the regions, where equipotentials from different planes 
touch each other, play the role of saddle points for such a network.
The energy separation between two delocalized states is of the order of 
the width of the Landau level, and the critical exponent of the 
localization length is $7/3$--the same as for an isolated layer.
                                                                           
\end{abstract}

\begin{multicols}{2}
\narrowtext

\section{Introduction}
\label{I}
In a recent paper \cite{SorMD}, S{\o}rensen  and MacDonald 
studied the integer quantum Hall effect in a double--layer 
system with uncorrelated disorder. They investigated numerically 
the localization properties of single--electron states for
two limiting cases: of a short--range and of a smooth disorder.
For a short--range disorder, the relevant parameter is 
$t/ \Gamma$, which is the ratio of the tunnel integral and 
the disorder--induced width of the Landau level. If this ratio is 
large,  the situation is transparent since localization 
of  states, belonging to  symmetric and antisymmetric
combinations of the size--quantization  wave functions,
occurs independently. Thus, one has two delocalized 
levels, located at the two well--separated maxima of the density of states, 
and spaced in energy by $2t$. 

When $t/ \Gamma \ll 1$, the density of states has only 
one maximum; the usage of the basis of symmetric and antisymmetric 
combinations is inadequate. However, it was demonstrated 
in the simulation of  Ref. \onlinecite{SorMD} (see also 
Ref. \onlinecite{OOK})
that the separation of delocalized states is close to $2t$ even
for $t/ \Gamma $ as small as $0.1$. In other words, 
the positions of delocalized states with uncorrelated 
disorder appear to be the 
same as in the case of correlated disorder. 

To study the case of a smooth disorder, the authors \cite{SorMD}
employed the network model proposed by 
Chalker and Coddington \cite{CC}, which was generalized 
to describe a double--layer system in a way similar to  that in 
Refs. \onlinecite{LC,LCK,CD}. In the Chalker--Coddington model, the 
delocalization results from the tunneling of an electron,
moving along equipotential lines, through the saddle points 
of the random potential. The randomness of the potential is included
by assuming the phase acquired by an electron traversing a link 
(equipotential line) to be random.
For a double--layer system  each link of a network carries two channels.
The authors tried several variants of incorporating the inter--layer 
tunneling through the coupling between the channels.
The results obtained were essentially the same. 
In contrast to the case of two spin--split levels with random mixing
\cite{LC,LCK,HAMG}, the authors observed the doubling of the 
correlation length exponent.
One of the main conclusions made  in Ref. \onlinecite{SorMD} 
on the basis of the simulation performed, is that 
there is no observable splitting in the positions 
of the delocalized states within the range of $t/ \Gamma$ studied,
which makes doubtful the applicability of the network model to the
double--layer systems.
Motivated by this observation, in the present paper we take a
microscopic approach to the problem.

In a strong magnetic field the structure of electronic states in 
the presence of a smooth potential is determined by its topography.
 The crucial difference between the cases
of correlated and uncorrelated disorder is that in the latter case
the equipotentials from the  different layers, may cross each other
after projection on the same plane. This is illustrated in Fig. 1.
As a result of such a crossing, there is a finite probability 
for an electron to change equipotential (and, correspondingly, the plane).
Obviously, this probability increases with increasing $t$.
We will show that the probability $\cal{P}$, for an electron 
to stay within the same equipotential  after crossing, is given by 
\begin{equation}
\label{P}
{\cal{P}} =  \exp \left(- \frac{2\pi t^2 l^2}
{|{\bf{v}}_1 \times {\bf{v}}_2| \hbar^2} \right)  ,
\end{equation}
where $l$ is the magnetic length and ${\bf{v}}_1$, ${\bf{v}}_2$ 
are the drift velocities of an electron along the first and the 
second equipotentials respectively.
These velocities are related to the values
${\mbox{\boldmath${\cal{E}}$}}_1$,
${\mbox{\boldmath${\cal{E}}$}}_2$ of 
the electric field in each of the planes at the point of crossing 
as: ${\bf{v}}_{1,2}= c({\mbox{\boldmath${\cal{E}}$}}_{1,2} 
\times  {\bf{B}})/B^2$, 
where $\bf{B}$ is the magnetic field perpendicular to the planes.
Then we can rewrite (\ref{P}) as ${\cal{P}} =  \newline
\exp(-2\pi t^2 /e^2 |{\mbox{\boldmath{${\cal{E}}$}}}_1 \times 
{\mbox{\boldmath${\cal{E}}$}}_2|l^2)$. This sets a relevant scale for $t$.
Indeed, the fields ${\cal{E}}_{1,2}$  can be estimated as 
$\Gamma / e R_c$, where $R_c$ is the correlation radius of a 
smooth potential.
\begin{figure}
\label{Fig.1}
\epsfysize=3.5in
\centerline{\epsffile{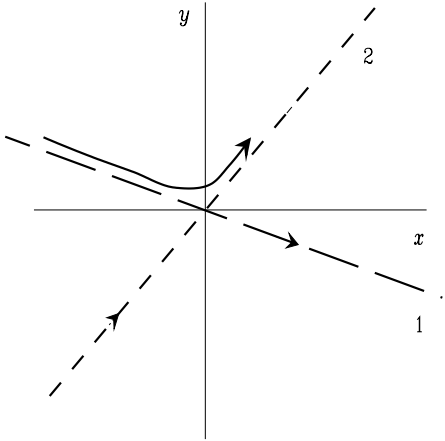}}
\caption{Crossing of equipotentials  
from the  first \newline (long--dashed lines) and the  second
(dashed lines) layer after projection on the same plane.}
\end{figure}
Thus we see, that the probability to retain equipotential  is close to 
$1$ if $t \ll \Gamma l/ R_c$. In the opposite case, $t \gg  \Gamma l/ R_c$,
practically each crossing results  in the change of equipotential.
It is very important that the crossover value $t \sim  \Gamma l/ R_c$
is much smaller than $\Gamma$ if the disorder is smooth.
This implies that, as $t$ gradually increases, the regime when the 
equipotential is switched with probability close to $1$ establishes
already at rather small values of $t$, where the density of states still
represents a single peak.
The applicability of the formula (\ref{P}) to the double--layer system
with uncorrelated random potential is limited by the condition 
$t \ll \Gamma$. This is because in the derivation of (\ref{P}) it 
was assumed that  electric fields ${\mbox{\boldmath{${\cal{E}}$}}}_1$,
${\mbox{\boldmath{${\cal{E}}$}}}_2$ 
are constant within the entire region where 
the ``interaction''  between equipotentials occurs.
One can see from the derivation presented below, that this region is
of the order of the magnetic length when 
$t \ll e {\cal{E}}_{1,2}l$, i.e. when  $\cal{P}$ is close to $1$.
However, in the opposite limit, $t \gg e {\cal{E}}_{1,2}l$, 
which corresponds to the strong coupling, the extension of 
the interaction region is of the order  of $t/e {\cal{E}}_{1,2}$
and is much larger than $l$. On the other hand, for Eq.(\ref{P}) 
to be relevant, this extension should be much smaller than $R_c$, 
which leads us to the condition $t \ll \Gamma$.
Note that for  $t \gg \Gamma$, the language of symmetric and 
antisymmetric states becomes adequate. Then, similarly to the case of the
short--range disorder, the density of states represents two 
peaks, each having a delocalized state in the center.

Thus, we have established that in a wide region 
$\Gamma l/R_c \ll t \ll \Gamma$, each crossing of equipotential lines 
from different planes leads to the change of the plane, 
in which the electron  moves. As a result, the saddle points \cite{CC}  of
the random potentials $V_1(\mbox{\boldmath$\rho$})$ and 
$V_2(\mbox{\boldmath$\rho$})$
in the planes, which played a
crucial role for delocalization at $t=0$, become irrelevant 
in this region (a typical saddle point would be bypassed due to switching of
equipotentials, see Fig. 2).
\begin{figure}
\label{Fig.2}
\epsfysize=3.3in
\centerline{\epsffile{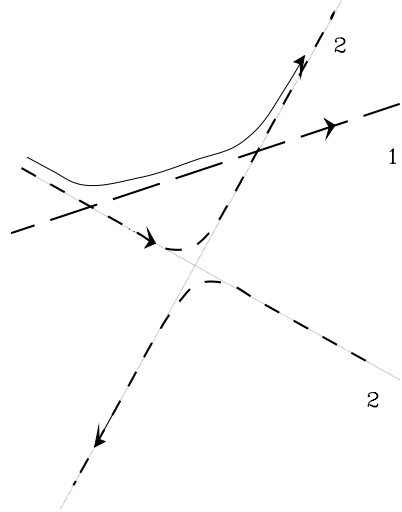}}
\caption{Bypassing of a saddle point due to switching of equipotentials
at the intersections.}
\end{figure}
It appears that within the region  $\Gamma l/R_c < t < \Gamma$, 
the delocalized states have completely different structure  
than at $t=0$.
Their energy positions, measured from the center of Landau level,
are $E=\pm E_c$, where $E_c$ is of the order 
of the width of the Landau level $\Gamma$ (see Fig. 3), and they 
are composed of alternating
pieces of compact loops, which are equipotentials 
$V_1(\mbox{\boldmath$\rho$})=E_c$ and $V_2(\mbox{\boldmath$\rho$})=E_c$
for one state and  
$V_1(\mbox{\boldmath$\rho$})=-E_c$ and $V_2(\mbox{\boldmath$\rho$})=-E_c$
for the other  state.
To justify this scenario, assume first that energy $E$ 
is deep in the tail of the Landau level.
Then the equipotentials 
$V_1(\mbox{\boldmath$\rho$})=E$ and $V_2(\mbox{\boldmath$\rho$})=E$,
being projected on the same plane,  represent  a set of isolated circles.
As $E$ moves up, the equipotentials, corresponding to different 
planes, start to overlap and form clusters, as it is shown in Fig. 4 .
It is important  that, due to switching of equipotentials at intersections,
the motion of an electron within a cluster occurs either 
inside the cluster or along  its boundary
(see Fig. 4). 
As $E$ further increases, 
the average size (correlation radius) of the  clusters grows and,
at some $E=E_c \sim \Gamma$, the classical percolation occurs. 
As the energy sweeps  through $E_c$, critical clusters first come close,
touch each other, and, finally,  merge.

To describe the corresponding change in the electronic  states, 
quantum mechanics becomes important.
We will show that there is a complete correspondence between a region
of  touching of equipotentials and a conventional saddle point.
Namely, the transmission coefficient, defined as a probability for an 
electron  to change the
cluster while passing  through the region of  touching,  
 has the same form as the transmission coefficient of a saddle 
point\cite{FH}
\begin{equation}
\label{TFH}
T=\frac{1}{1+\exp\left[\frac{\pi (E-\tilde{E}_c)}{\Delta}\right]},
\end{equation} 
where $\tilde{E}_c$ is the ``height'' of the effective saddle point 
and it differs from the value of potential at the point of touching 
by an amount of the order of $t$.
For the characteristic energy $\Delta$ we derive below an estimate:
$\Delta \sim \Gamma^{3/2}l^2/t^{1/2}R_c^2$. It is easy to see 
that in the region $\Gamma l/R_c < t < \Gamma$ one has 
$\Delta \ll \Gamma$.

Thus we  see, that the description of delocalization at $E \approx \pm E_c$
reduces to a   single--channel network model with 
critical clusters, playing the role of links, and the regions of  touching
acting as nodes. 
\begin{figure}
\label{Fig.3}
\epsfysize=3.3in
\centerline{\epsffile{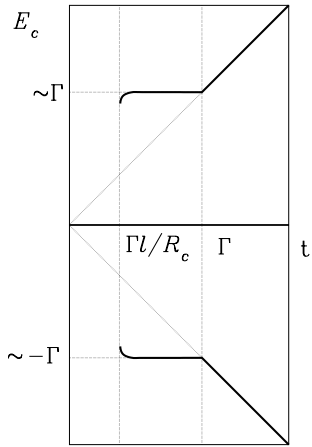}}
\caption{Schematic plot of the energy positions of delocalized states in
a double--layer system at different values  of interlayer coupling.}
\end{figure}
This implies that the
delocalized states  have the same critical exponent of the localization 
length \cite{CC,LWK,HK,HB,WLW}, $\kappa= 7/3$, as in an isolated layer.
The magnitude of the localization length can be estimated as follows.
Quantum mechanical description of the transmission through the effective
saddle point becomes important when $|E-E_c| < \Delta$. The size of the 
``unit cell'' of the network for $E=E_c \pm \Delta$ is 
$\sim R_c (\Gamma/\Delta)^{4/3}$, where $4/3$ is the critical exponent 
for the classical percolation. 
Then we have, 
\begin{equation}
\label{Corr_radius}
\xi \sim R_c \left(\frac{\Gamma}{\Delta}\right)^{\frac{4}{3}}
\left(\frac{\Delta}{E-E_c}\right)^{\frac{7}{3}}
\sim \frac{l^2}{R_c} 
\frac{\Gamma^{\frac{17}{6}}}{t^{\frac{1}{2}} (E-E_c)^{\frac{7}{3}}}.
\end{equation}

As $t$ becomes smaller than $\Gamma l/R_c$, the above picture of electronic 
states is not valid anymore. A typical crossing of equipotentials would 
not cause the change  of  the plane for the electron motion. 
We can present only a plausible argument about the evolution of 
the positions of delocalized states in this limit. Note that, although a
typical crossing is not efficient for $t < \Gamma l/R_c$, the regions 
of touching of equipotentials from different planes still act as
saddle points. It will be shown below that the corresponding condition on 
$t$ is $t > \Gamma (l/R_c)^{4/3}$. Thus, in the domain 
$\Gamma l/R_c > t > \Gamma (l/R_c)^{4/3}$, these effective 
saddle points would couple electronic states belonging to closed 
equipotentials from different planes. As a result, the delocalized states would
occur at energies $\pm E_c^{(1)}$, at which  the size of a closed 
equipotential, $R_c (\Gamma/E_c^{(1)})^{4/3}$, is big enough to have $\sim 1$
effective saddle point somewhere on its perimeter. It is obvious that 
$ E_c^{(1)}\ll E_c$. Then the problem again reduces to a single--channel
network model with the cells of the network being closed
equipotentials, alternatingly, from the first and from the second layer,
and the nodes being the effective saddle points.
\begin{figure}
\label{Fig.4}
\epsfysize=3.3in
\centerline{\epsffile{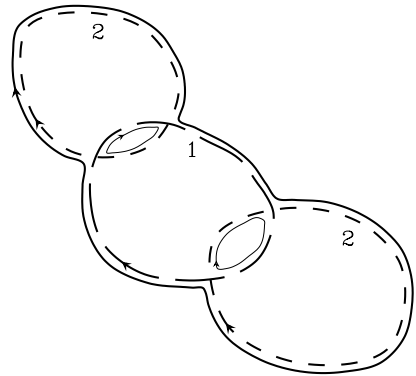}}
\caption{A cluster formed by equipotentials from different layers
after projection on the same plane. Full lines show the resulting trajectories
of the electron motion.}
\end{figure}

To estimate the magnitude of $E_c^{(1)}$ note that the perimeter of
critical equipotential scales with energy as\cite{aha}: ${\cal L}(E)
\sim R_c(\Gamma/E)^{7/3}$. The probability that two equipotentials from
different layers would come close and form an effective saddle point 
with height within the interval $(E-\Delta, E+\Delta)$ can be estimated as 
$\Delta/\Gamma\sim (\Gamma/t)^{1/2}(l/R_c)^2$. Remember that $\Delta$ is
the energy scale of the transmission coefficient in Eq. (\ref{TFH}).
Then the condition to find $E_c^{(1)}$ can be written as:
$({\cal L}(E_c^{(1)})/R_c)(\Delta/\Gamma) \sim 1$, and it yields
$E_c^{(1)}\sim \Gamma (\Gamma/t)^{3/14}(l/R_c)^{6/7}$.

In the next two sections we derive the basic equations, 
(\ref{P}) and (\ref{TFH}). Section \ref{Con} concludes the paper.

\section{Transitions between  intersecting equipotentials}
\label{T}
Suppose that two equipotentials from different planes, 
corresponding to the energy E, measured from the center of the 
Landau level, intersect at $\rho=0$.
The potentials $V_1(\mbox{\boldmath{$\rho$}})$,
$V_2(\mbox{\boldmath{$\rho$}})$ behave  near $\rho=0$ as 
\begin{equation}
\label{V}
V_1(\mbox{\boldmath{$\rho$}})=E+
e {\mbox{\boldmath{${\cal{E}}$}}}_1 \cdot {\mbox{\boldmath{$\rho$}}};
\quad
V_2(\mbox{\boldmath{$\rho$}})=E+
e {\mbox{\boldmath{${\cal{E}}$}}}_2 \cdot {\mbox{\boldmath{$\rho$}}}
\end{equation} 
Then the amplitudes $\Psi_1$ and $\Psi_2$, to find an electron respectively
in the first and in the second plane, satisfy the following system of 
equations
\begin{eqnarray}
\label{Sch01}
\left[\frac{1}{2m} \left(\hat{\bf{p}}+\frac{e}{c}{\bf{A}} \right)^2
+ e {\mbox{\boldmath{${\cal{E}}$}}}_1 \cdot {\mbox{\boldmath{$\rho$}}} - 
\hbar \omega \left(n+\frac{1}{2} \right) \right]
\Psi_1 \nonumber \\ = t \Psi_2, \\
\label{Sch02}
\left[\frac{1}{2m} \left(\hat{\bf{p}}+\frac{e}{c}{\bf{A}} \right)^2
+ e {\mbox{\boldmath{${\cal{E}}$}}}_2 \cdot {\mbox{\boldmath{$\rho$}}}-
\hbar \omega \left(n+\frac{1}{2} \right) \right]
\Psi_2 \nonumber \\  = t \Psi_1.
\end{eqnarray}
We chose a symmetric gauge ${\bf{A}}=B(-y,x,0)/2$
and perform a  transformation to the new variables, defined as
\begin{eqnarray}
\label{GCC}
x&=&l(X-s), 
\qquad \qquad
y=-i l \left(\frac{\partial}{\partial X}+
\frac{\partial}{\partial s} \right), \\
\frac{\partial}{\partial x}&=&\frac{1}{2l} 
\left(\frac{\partial}{\partial X}- \frac{\partial}{\partial s} \right),
\quad  
\frac{\partial}{\partial y}= -\frac{i}{2l} \left(X+s \right).
\end{eqnarray}
This transformations was  previously used by Fertig and Halperin \cite{FH}
to describe the transmission through a saddle point. It allows to separate 
effectively the cyclotron motion ($s$--coordinate), and the motion of the
guiding center ($X$--coordinate).
With new variables, the system (\ref{Sch01}),(\ref{Sch02}) reads 
\begin{eqnarray}
\label{Sch1}
&&\bigg(-\frac{\partial^2}{\partial s^2}-
i {\beta}_{1y}\frac{\partial}{\partial s}+
s^2 -{\beta}_{1x}s \bigg)\Psi_1+  \nonumber \\ 
&&\bigg({\beta}_{1x} X -
i {\beta}_{1y}
\frac{\partial}{\partial X}- 2n-1 \bigg) \Psi_1=
\frac{2t}{\hbar \omega} \Psi_2 
\end{eqnarray}
\begin{eqnarray}
\label{Sch2}
&&\bigg(-\frac{\partial^2}{\partial s^2}-
i {\beta}_{2y}\frac{\partial}{\partial s}+
s^2 - {\beta}_{2x}s \bigg)\Psi_2+  \nonumber \\ 
&& \bigg({\beta}_{2x} X -
i {\beta}_{2y}
\frac{\partial}{\partial X}- 2n-1 \bigg) \Psi_2=
\frac{2t}{\hbar \omega} \Psi_1,  
\end{eqnarray}
where we have introduced the notations:
$\beta_{1,2x}=2 e {\cal{E}}_{1,2x}l/\hbar \omega$,
$\beta_{1,2y}=2 e {\cal{E}}_{1,2y}l/\hbar \omega $  
and  $\omega=eB/mc$, is the cyclotron frequency.
The term proportional to $\partial / \partial s$ can be eliminated
by the following substitution
\begin{equation}
\label{T1}
\Psi_{1,2}=\exp\left[-i \frac{{\beta}_{1,2y}}{2}
\left(s- \frac{{\beta}_{1,2x}}{2}\right) \right]
\Phi_{1,2} \left( s- \frac{{\beta}_{1,2x}}{2}\right).
\end{equation}
As a result the equations (\ref{Sch1}),(\ref{Sch2}) acquire a separable form
\begin{eqnarray}
\label{Sch11}
\lefteqn{\left[-\frac{\partial^2}{\partial s^2}+
\left(s- \frac{{\beta}_{1x}}{2}\right)^2 \right] \Phi_1 +
\left({\beta}_{1x} X -
i {\beta}_{1y}
\frac{\partial}{\partial X} \right) \Phi_1-} \nonumber \\
&&\left(2n+1+{\beta}_1^2 \right) \Phi_1=  \nonumber \\
&& \frac{2t}{\hbar \omega} \;
\exp  \left\{i \frac{{\beta}_{1y}-{\beta}_{2y}}{2}s 
-i \frac{{\beta}_{1x}{\beta}_{1y}-
{\beta}_{2x}{\beta}_{2y}}{4}
\right\}  \Phi_2  ,
\end{eqnarray}
\begin{eqnarray}
\label{Sch22}
\lefteqn{\left[-\frac{\partial^2}{\partial s^2}+
\left(s- \frac{{\beta}_{2x}}{2}\right)^2 \right] \Phi_2  +
\left({\beta}_{2x} X -
i {\beta}_{2y}
\frac{\partial}{\partial X} \right) \Phi_2 -} \nonumber \\
&&\left(2n+1+{\beta}_2^2 \right) \Phi_2=  \nonumber \\
&& \frac{2t}{\hbar \omega} \;
\exp \left\{-i \frac{{\beta}_{1y}-{\beta}_{2y}}{2}s 
+i \frac{{\beta}_{1x}{\beta}_{1y}-
{\beta}_{2x}{\beta}_{2y}}{4}
\right\}  \Phi_1 ,
\end{eqnarray}
where ${\beta}_1^2={\beta}_{1x}^2+{\beta}_{1y}^2$
and ${\beta}_2^2={\beta}_{2x}^2+{\beta}_{2y}^2$.
Since $X$ enters linearly in both equations, 
we will search  for solutions in the following form:
\begin{eqnarray}
\label{T2}
\label{Phi1}
\Phi_1&=& F_n(X) \phi_n \left(s- \frac{{\beta}_{1x}}{2} \right), \\
\label{Phi2}
\Phi_2&=& G_n(X) \phi_n \left(s- \frac{{\beta}_{2x}}{2} \right),
\end{eqnarray}
where $\phi_n$ are the oscillator eigenfunctions.
Then the system of equations for the  coefficients $F_n, G_n$, 
describing  the motion of the guiding center, takes the form
\begin{eqnarray}
\label{Sch111}
\left(-i {\beta}_{1y} \frac{\partial}{\partial X}+
{\beta}_{1x} X \right) F_n - {\beta}_1^2 F_n= 
\frac{2t}{\hbar \omega} G_n, \\
\label{Sch222}
\left(-i {\beta}_{2y} \frac{\partial}{\partial X}+
{\beta}_{2x} X \right) G_n - {\beta}_2^2 G_n= 
\frac{2t}{\hbar \omega} F_n,
\end{eqnarray}
We have set unity the overlap integral of the functions $\phi_n$ 
centered at $\beta_{1x}/2$ and $\beta_{2x}/2$.
Indeed, one has $ {\beta}_{1x}-{\beta}_{2x} \sim 
e {\cal{E}}_{1,2}l/ \hbar \omega \ll 1$.
In fact, the system (\ref{Sch111}),(\ref{Sch222}) is equivalent 
to the system describing non--adiabatic transitions between 
the crossing energy levels in molecules. 
This problem was first considered more than 
60 years ago \cite{Z}.  Up to a phase factor, the 
solution of the system (\ref{Sch111}), (\ref{Sch222}) 
can be expressed in terms of parabolic 
cylinder functions \cite{AE}, $D_{\nu}(\pm \overline{X} 
\sqrt{2}e^{i\pi /4})$, with $\nu$ and $ \overline{X}$ 
given by the following formulas:
\begin{eqnarray}
\label{nu}
\nu = -i \frac{t^2}{e^2 l^2 |{\mbox{\boldmath${\cal{E}}$}} _1 \times 
{\mbox{\boldmath${\cal{E}}$}}_2|} = 
-i \frac{t^2 l^2}{|{\bf{v}}_1 \times {\bf{v}}_2| \hbar^2} , \\
\label{Xbar}
 \overline{X}= \sqrt{\frac{2 {\cal{E}}_{1y}
{\cal{E}}_{2y}}
{|{\mbox{\boldmath{${\cal{E}}$}}}_1 \times 
{\mbox{\boldmath${\cal{E}}$}}_2|}} \left[ X-
\frac{\hbar \omega}{2} \frac{\varepsilon_n 
({\cal{E}}_{1y}-
{\cal{E}}_{2y})}
{e l |{\mbox{\boldmath{${\cal{E}}$}}}_1 \times 
{\mbox{\boldmath${\cal{E}}$}}_2|} \right] .
\end{eqnarray}
Using the asymptotics of the $D$--functions,
\begin{eqnarray}
\label{AS1}
D_{\nu}(X) \sim  X^{\nu}e^{-\frac{1}{4} X^2} &&, \nonumber \\
  |X| \rightarrow \infty, &&
 \ \ \left( - \frac{3 \pi}{4}< argX < \frac{3 \pi}{4} \right) , 
\end{eqnarray}
\begin{eqnarray}
\label{AS2}
D_{\nu}(X) \sim X^{\nu}e^{-\frac{1}{4} X^2}-
\frac {(2 \pi)^{1/2}}{\Gamma (- \nu)}e^{- i \pi \nu }X^{- \nu -1}
e^{\frac{1}{4} X^2}, \nonumber \\
  |X| \rightarrow \infty, \ \ 
\left(- \frac{5 \pi}{4}< argX < -\frac{\pi}{4}\right), 
\end{eqnarray}
one can see that the right behavior at $X \rightarrow + \infty$
for the function $F_n(X)$ (no reflected wave), is insured 
by the following choice:
\begin{equation}
\label{F}
F_n(X) \propto  D_{\nu} \left(\pm \overline{X} 
\sqrt{2}e^{i\pi /4} \right).
\end{equation}
The asymptotics for $F$ at $\overline{X} \rightarrow \infty$
and $\overline{X} \rightarrow -\infty$, differ by a factor 
$\exp(i \pi \nu)$. With $\nu$ given by (\ref{nu}), the probability 
to retain the equipotential after crossing, 
${\cal{P}}= |F(-\infty)|^2/|F(\infty)|^2$,
takes the form Eq. (\ref{P}).

As it was mentioned in the Introduction, the characteristic spatial 
scale of the interaction region depends of the value of $t$.
If $t \ll e {\cal{E}}_{1,2} l$, we have $|\nu| \ll 1$ 
and the characteristic $X$ in (\ref{F}) is of the order of $1$.
This means that the spatial scale for  the interaction  is $l$.
In the opposite case, $t \gg e {\cal{E}}_{1,2} l$, the parameter 
$\nu$ is large, $|\nu| \gg 1$, and the product 
$X^{\nu}\exp(-X^2/2)$ in (\ref{F}) sets a scale for $X$:
$X \sim |\nu|^{1/2} \sim t/e {\cal{E}}_{1,2} l$, which leads 
to the characteristic scale $\sim t/e {\cal{E}}_{1,2}$ for 
the interaction region.

\section{The effective saddle point}
\label{SP}
Assume for concreteness that two equipotentials that nearly touch each 
other, have their origin in two displaced potential minima in 
each layer (Fig. 5)
\begin{eqnarray}
\label{ESP1}
V_1(x,y)=\frac{m\Omega_1^2}{2} \left[(y-y_1)^2+x^2 \right]+V_{10} \nonumber \\
\approx \frac{m\Omega_1^2}{2} \left[x^2-2y_1 y+y_1^2 \right]+V_{10}, \\
\label{ESP2}
V_2(x,y)=\frac{m\Omega_2^2}{2} \left[(y+y_2)^2+x^2 \right]+V_{20} \nonumber \\
\approx \frac{m\Omega_2^2}{2} \left[x^2+2y_2 y+y_2^2 \right]+V_{20},
\end{eqnarray}
where $(y_1,0)$ and $(-y_2,0)$ are the positions of the minima,
$V_{10},V_{20}$ are the heights and $\Omega_1, \Omega_1$ are the curvatures.
We neglect the terms $\sim y^2$ in (\ref{ESP1}), (\ref{ESP2}) since 
the relevant $y$ appears to be small.
\begin{figure}
\label{Fig.5}
\epsfysize=3.3in
\centerline{\epsffile{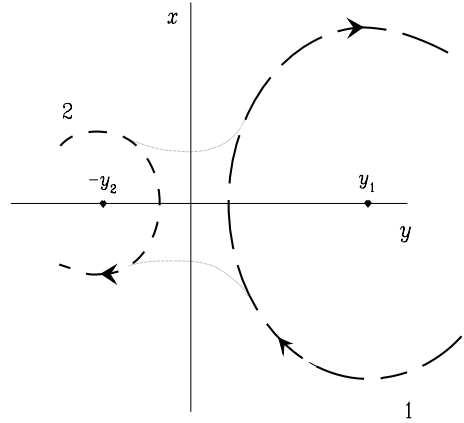}}
\caption{Effective saddle point formed by equipotentials from different
 layers after projection
on the same plane.}
\end{figure}
The condition that two equipotentials come close to each other
at $x=0, y=0$, can be expressed as 
\begin{equation}
\label{DSP}
\frac{m\Omega_1^2y_1^2}{2}+V_{10} \approx 
\frac{m\Omega_2^2y_2^2}{2} + V_{20} \approx E
\end{equation}
Then the system of equations for the amplitudes $\Psi_1, \Psi_2$,
similar to (\ref{Sch01}), (\ref{Sch02}), can be written as
\begin{eqnarray}
\label{SSch1}
\left[ \frac{\hat{p}_x^2}{2 m}+ \frac{m(\omega^2+\Omega_1^2)}{2}
\left(x-\frac{\omega}{\omega^2+
\Omega_1^2} \frac{\hat{p}_y}{m} \right)^2 \right] \Psi_1 + \nonumber \\
\left(\frac{\Omega_1^2}{2m \omega^2 } \hat{p}_y^2 - 
m \Omega_1^2 y_1 y \right) \Psi_1 -  \nonumber \\
 \left[E-\hbar \omega \left(n+\frac{1}{2}\right)-
\frac{m\Omega_1^2y_1^2}{2}-V_{10} \right] \Psi_1 
= t \Psi_2 , 
\end{eqnarray}
\begin{eqnarray}
\label{SSch2}
\left[ \frac{\hat{p}_x^2}{2 m}+ \frac{m(\omega^2+\Omega_2^2)}{2}
\left(x-\frac{\omega}{\omega^2+
\Omega_2^2} \frac{\hat{p}_y}{m} \right)^2  \right] \Psi_2 + \nonumber \\
\left(\frac{\Omega_2^2}{2m \omega^2 } \hat{p}_y^2 +
m \Omega_2^2 y_2 y \right) \Psi_2 - \nonumber \\
\left[E-\hbar \omega \left(n+\frac{1}{2}\right)- 
\frac{m\Omega_2^2y_1^2}{2}-V_{20} \right] \Psi_2
= t \Psi_1.
\end{eqnarray}
We will search for a solution in the following form:
\begin{eqnarray}
\label{ST1}
\Psi_1&=& \int dk A(k)e^{iky}\phi_n \left(x-\frac{\omega^2}{\omega^2+
\Omega_1^2} \l^2 k\right), \\
\label{ST2}
\Psi_2&=& \int dk B(k)e^{iky}\phi_n \left(x-\frac{\omega^2}{\omega^2+
\Omega_2^2} \l^2 k\right).
\end{eqnarray}
Upon substituting (\ref{ST1}), (\ref{ST2}) into  (\ref{SSch1}), (\ref{SSch2}),
we get a system of equation for the functions $A$ and $B$ 
\begin{eqnarray}
\label{SSch11}
-i m \Omega_1^2 y_1 \frac{\partial A}{\partial k}+a(k) A(k)&=& t c(k) B(k), \\
\label{SSch22}
+i m \Omega_2^2 y_2 \frac{\partial B}{\partial k}+b(k) B(k)&=& t c(k) A(k),
\end{eqnarray}
where $c(k)$ is the overlap integral
\begin{eqnarray}
\label{c}
c(k)&&= \nonumber \\ 
\int&& dx  \phi_n  \left(x-\frac{\omega^2 \l^2 k }{\omega^2+
\Omega_1^2} \right) \phi_n \left(x-\frac{\omega^2 \l^2 k }{\omega^2+
\Omega_2^2} \right),
\end{eqnarray}
and the coefficients $a(k)$ and  $b(k)$ are defined as  
\begin{eqnarray}
\label{a}
a(k) &=& \frac{m \Omega_1^2 l^4 k^2}{2}- 
\left(E-\frac{m \Omega_1^2 y_1^2}{2}-V_{10} \right), \\
\label{b}
b(k) &=& \frac{m \Omega_2^2 l^4 k^2}{2}- 
\left(E-\frac{m \Omega_2^2 y_2^2}{2}-V_{20} \right).
\end{eqnarray}
It can be easily seen that $c(k)=1$ for $\Omega_1=\Omega_2$
and the correction in the case when the two frequencies are different,
is proportional to $(\Omega_1^2-\Omega_2^2)^2/ \omega^4$. Since the  
the random potential is smooth, we can neglect this correction 
and set $c(k)=1$. The system (\ref{SSch11}), (\ref{SSch22}) 
can be  reduced to a single 
second--order differntial equation, say, for $A(k)$
\begin{eqnarray}
\label{A}
\frac{d^2A}{dk^2}+i \left(\frac{a}{m \Omega_1^2 y_1}-
\frac{b}{m \Omega_2^2 y_2}\right) \frac{dA}{dk}+ \nonumber \\
\left[\frac{ab-t^2}{m^2 \Omega_1^2  \Omega_2^2 y_1 y_2}+
\frac{i}{m \Omega_1^2 y_1} \frac{da}{dk} \right] A=0.
\end{eqnarray}
The term with first derivative  can be eliminated by the 
following substitution
\begin{equation}
\label{TA}
A(k)= 
\exp \left[-\frac{i}{2} \int_{-\infty}^k dk^{\prime} 
\left(\frac{a(k^{\prime})}{m \Omega_1^2 y_1}-
\frac{b(k^{\prime})}{m \Omega_2^2 y_2}\right) \right] {\cal{A}}(k),
\end{equation}
after which Eq. (\ref{A}) takes the form
\begin{eqnarray}
\label{Arond}
\frac{d^2 {\cal{A}}}{dk^2}+
\bigg[\frac{1}{4}\left(\frac{a}{m \Omega_1^2 y_1}+
\frac{b}{m \Omega_2^2 y_2}\right)^2 &&-  \nonumber \\
 \frac{t^2}{m^2 \Omega_1^2 \Omega_2^2 y_1 y_2}+
\frac{i l^4 k}{2} \left(\frac{1}{y_1}+\frac{1}{y_2} \right) 
\bigg]&& {\cal{A}}= 0 .
\end{eqnarray}
It is convenient to introduce the following notations: 
\begin{eqnarray}
\label{yomega0 }
\frac{2}{y_0}= \frac{1}{y_1}+\frac{1}{y_2}, \quad 
\Omega_0^2 =
\Omega_1 \Omega_2 \frac{\sqrt{y_1 y_2}}{y_0},\\
\label{energy0}
\varepsilon_0=
\frac{\Omega_0^2 y_0}{2  \Omega_1^2 y_1} 
\left[E-\frac{m \Omega_1^2 y_1^2}{2}-V_{10} \right] + \nonumber \\
\frac{\Omega_0^2 y_0 }{2 \Omega_2^2 y_2} 
\left[E-\frac{m \Omega_2^2 y_2^2}{2}-V_{20} \right] .
\end{eqnarray}
Using the definitions of $a$ and $b$, the equation (\ref{Arond}) becomes
\begin{eqnarray}
\label{Arond1}
&&\frac{d^2 {\cal{A}}}{dk^2}+ \nonumber \\
\bigg[&&\bigg(\frac{{\varepsilon}_0}{m \Omega_0^2 
y_0}-
\frac{l^4 k^2}{2 y_0} \bigg)^2-
\left(\frac{t}{m \Omega_0^2 y_0} \right)^2+
\frac{i l^4 k}{y_0} 
\bigg] {\cal{A}}= 0 .
\end{eqnarray}
Eq. (\ref{Arond1}) has the form of the Schr\"{o}dinger equation with 
a complex ``potential energy''.
However, not all the terms in the ``potential energy'' are relevant.
This becomes obvious if we introduce the following rescaling 
of the argument 
\begin{equation}
\label{Res}
k=\frac{1}{l} \left(\frac{m \Omega_0^2 y_0^2}
{t} \right)^{\frac{1}{4}} z.
\end{equation}
Then the equation (\ref{Arond1}) takes the form 
\begin{equation}
\label{Arond11}
\frac{d^2 {\cal{A}}}{dz^2}+
\left[-z^2 \frac{{\varepsilon}_0}{t} + \frac{{\varepsilon}_0^2-t^2}
{t^2 \alpha^{2}} 
+ \frac{1}{4} z^4 
\alpha^2
+i z \alpha
\right] {\cal{A}}= 0 ,
\end{equation}
where we have introduced the dimensionless parameter 
\begin{equation}
\label{alfa}
\alpha= \frac{l}{y_0} \left(\frac{m \Omega_0^2 y_0^2}{t} \right)^{\frac{3}{4}}.
\end{equation}
This parameter can be estimated as follows. The typical value of $y_0$
is $\sim R_c$; the curvature $\Omega_0$ can be found from the condition\cite{psh}:
$m\Omega_0^2 R_c^2 \sim \Gamma$. Then we have 
$\alpha \sim (l/R_c)(\Gamma/t)^{3/4}$. But within the domain
we are interested in, $t$ is larger than 
$\Gamma l / R_c$. Then we get $\alpha < (l/R_c)^{1/4} \ll 1$.
This allows to drop the last two terms in the potential energy. 
It can be also seen that the effective saddle point corresponds to 
$|\varepsilon_0+t| \ll t$. 
Indeed, under this condition Eq. (\ref{Arond11}) takes the form 
\begin{equation}
\label{Arond111}
\frac{d^2 {\cal{A}}}{dz^2}+
\left[z^2 - 2 \frac{ \left(\varepsilon_0+ t \right)}{t \alpha^2} 
\right] {\cal{A}}= 0 .
\end{equation}
which is the equation, describing the scattering 
from the inverted parabolic potential. The expression for the 
transmition coefficient for this potential is 
well-known \cite{LL}
\begin{equation}
\label{TC}
T(E) = \frac{1}{1+
\exp\left[\frac{2 \pi (\varepsilon_0 + t)}{t \alpha^2}\right]}.
\end{equation}
We see that the characteristic energy scale for the change of the 
transmission coefficient is
\begin{equation}
\label{Delta}
\Delta \sim  t \alpha^2 \sim \frac{\Gamma^{\frac{3}{2}}}
{t^{\frac{1}{2}}}
\left(\frac{l}{R_c}\right)^2.
\end{equation}
Using the above estimates, the condition  $\alpha \ll 1$,
which guarantees that the region of nearly touching 
of two equipotentials acts as a saddle point, can be rewritten 
as $t \gg \Gamma (l/R_c)^{4/3}$.
Note that $\Delta$ {\em decreases} with increasing $t$, reflecting the
fact that the larger is $t$, the closer should equipotentials approach
each other in order to form the effective saddle point.

\section{Conclusion}
\label{Con} 

The main result of the present paper is that for a double--layer system 
with smooth uncorrelated disorder, the position of delocalized states 
as a function of  coupling between the layers, has a wide plateau (Fig. 3). 
This is in contrast to the case when the disorder in both layers is 
strictly the same. In the latter case, the energy separation between
the localized states is just $2t$.
Note that the prediction we make can, in principle, be tested experimentally 
on a {\it single} sample, since, as it was shown in Ref. \onlinecite{HMD}, 
the tunnel integral $t$ can be effectively 
tuned (suppressed) by applying a parallel magnetic field.

When the authors of Ref. \onlinecite{SorMD} questioned the applicability
of the network model to double--layer systems, they implied that 
the network remains the same as the coupling between layers changes. 
Our considerations shows that a single channel network model applies 
within a wide range of $t$, but the  {\it network itself} experiences 
restructuring with $t$.
In particular, while at $t=0$ the nodes are the saddle points 
of the intralayer potential, at large enough coupling the nodes are 
the points where equipotentials from different layers touch each other.

Experimentally it might be hard to realize both,  smooth disorder within 
each layer and the absence of correlation between the layers. 
However, for our arguments to apply, it is sufficient  that the 
correlation between the layers is not absolute.
In other words, it should be allowed for two equipotentials, corresponding 
to the same energy, to be displaced by more than a magnetic length. 
Then our picture remains valid, but the correlation would lead 
to the shift of the percolation threshold 
to an energy much closer to the center of the Landau level, than in the case
of uncorrelated disorder. This implies that the ``splitting'' of the 
delocalized states in the regime when the coupling between equipotentials 
is already strong would still remain much smaller than $\Gamma$, i.e. 
the plateau shown in Fig. 3 would become lower, and hence, narrower.

In the paper we demonstrate that when the tunnel integral exceeds 
the characteristic value $\Gamma l/R_c$, the critical exponent, 
$\kappa$, of the localization length is the same as for $t=0$.
We also argue that the description based on a single--channel network model 
(and thus $\kappa =7/3$) is applicable within the interval 
$\Gamma l/R_c > t > \Gamma(l/R_c)^{4/3}$.
Therefore if the doubling of $\kappa$ established in Ref. \onlinecite{SorMD}
occurs, this may happen only for $t< \Gamma(l/R_c)^{4/3}$.
However, it seems more likely that the description in terms of a
single--channel network model with neighboring cells belonging to
different layers applies even at very small $t$. The reason why we 
anticipate this is the following. The solution of the model problem 
in Sect. \ref{SP} shows that if two minima of the random potential in different
layers are located anomalously close to each other ($y_1, y_2$ in 
(\ref{ESP1}),(\ref{ESP2})) are anomalously small), then the formation of the
effective saddle point becomes possible {\it even at very small} $t$.
Certainly, the smaller is $t$, more sparse these saddle points are.
At $t=0$ the localization length increases in each layer as $\xi(E)\propto 
E^{-7/3}$, and each localized state consists of many cells of the 
intralayer network, separated by saddle points. 
Then, if $t$ is finite and very small, at
some $E=E_c^{(2)}$ there will be $\sim 1$ effective saddle point per perimeter
of  a 
localized state in each layer. At this energy (as well as at $E=-E_c^{(2)}$)
 the states in two layers
would form a new network with a much larger  unit 
cell $\sim \xi (E_c^{(2)})$. Then  the localization length would
behave as $(E-E_c^{(2)})^{-7/3}$ with a prefactor  much bigger
than that for a single layer. Certainly, this scenario is only hypothetical.  

Note that there is a significant difference between our picture and the one 
outlined by S{\o}rensen and MacDonald \cite{SorMD}. The line of argument in 
Ref. \onlinecite{SorMD} is as follows. In the absence of a disorder 
an electron  residing initially, say, in the first layer, 
would oscillate between the layers with a period $\tau=2 \pi \hbar/t$.
When the disorder is present, the electron drifts within the layer 
with velocity $v_1$. It was assumed  in  Ref. \onlinecite{SorMD} that
the change of the layers would most probably occur after an electron 
travels the distance $l_{dr}=v_1 \tau = 2 \pi \hbar v_1/t \sim 
\Gamma l^2/ t R_c$.
In our picture the possibility for an electron in the first layer to tunnel
depends on the actual topography of the random potential in the second layer.
In contrast to Ref. \onlinecite{SorMD}, in our picture
 the change of the layers occurs
locally, at the intersections of equipotentials with the same energy.
At the same time in the regime of strong coupling, $t > \Gamma l/R_c$, that
we considered, the distance (in the vicinity of an intersection),
 over which the change of the layers takes place, is $\sim t R_c /\Gamma$ and
it is much larger than $l_{dr}$.

Our picture also differs from that by Laikhtman and Menashe\cite{LM}. 
Similarly to Ref. \onlinecite{SorMD}, they assume that the process of changing
layers, during the drift along the equipotential, occurs homogeneously, but   
they get a different estimate for the characteristic length of travel within
a given layer. Their estimate, $R_c^2 \Gamma/lt$, is larger than $t R_c /\Gamma$
and, correspondingly, larger than $l_{dr}$. 

\acknowledgments
The authors are grateful to T. V. Shahbazyan for reading the manuscript
and valuable remarks.
One of the authors (M. E. R.) is grateful to E. S{\o}rensen for 
a useful discussion. He is also grateful to the Aspen Center for Physics
for hospitality.

\end{multicols}

\end{document}